\documentclass[aps,prb,twocolumn,groupedaddress,showpacs,showkeys]{revtex4}

\usepackage{graphicx}

\begin{document}


\title{Force-matched embedded-atom method potential for niobium}


\author{Michael R. Fellinger}
\email[Correspondence author: ]{mfelling@mps.ohio-state.edu}
\affiliation{Department of Physics, The Ohio State University, Columbus, Ohio
43210, USA}
\author{Hyoungki Park}
\affiliation{Department of Physics, The Ohio State University, Columbus, Ohio
43210, USA}
\author{John W. Wilkins}
\affiliation{Department of Physics, The Ohio State University, Columbus, Ohio
43210, USA}


\date{\today}

\begin{abstract} Large-scale simulations of plastic deformation and
phase transformations in alloys require reliable classical interatomic
potentials. We construct an embedded-atom method potential for niobium
as the first step in alloy potential development. Optimization of the
potential parameters to a well-converged set of density-functional
theory (DFT) forces, energies, and stresses produces a reliable and
transferable potential for molecular dynamics simulations. The potential 
accurately describes properties related to the fitting data, and also produces 
excellent results for quantities outside the fitting range. Structural and 
elastic properties, defect energetics, and thermal behavior compare well with 
DFT results and experimental data, e.g., DFT surface energies are reproduced 
with less than 4\% error, generalized stacking-fault energies differ from
DFT values by less than 15\%, and the melting temperature is within 2\%
of the experimental value.

\end{abstract}


\maketitle


\section{\label{sec:intro} Introduction}

Niobium's low density, high melting temperature, and biocompatibility make it 
an attractive material for alloy design. Nb alloys are promising candidate 
materials for a wide variety technological applications. Multifunctional 
Ti-based ``gum metal'' alloys with substantial Nb concentrations exhibit 
remarkable properties and unique deformation behavior~\cite{saito03}. Attempts 
to increase the operating temperatures, and hence efficiencies, of turbine 
engines have prompted interest in designing Nb-based 
superalloys~\cite{ghosh07}. Non-toxic Ti-Nb based shape-memory alloys offer an 
alternative to Ti-Ni alloys for biomedical applications~\cite{kim05}. Accurate
and efficient computational models will aide in the microscopic understanding 
of deformation and transformation processes in these materials.

Advances in computational hardware and algorithms allow application of 
first-principles methods to increasingly complex problems. However, there 
remain calculations beyond the realm of {\it ab initio} methods. Meaningful 
simulations of processes involving long-ranged strain fields or long-wavelength
fluctuations require large numbers of atoms. Therefore, methods must be 
developed that scale more favorably with system size than first-principles 
calculations, while retaining a high degree of accuracy. The computational cost
for simulations based on short-ranged classical potentials scales linearly with
system size, allowing routine simulations of millions of atoms. However, the 
potentials must be carefully constructed and thoroughly tested to ensure that 
they yield reliable results. We construct a classical potential for large-scale
Nb simulations, and subsequent incorporation into potentials for alloys 
including Ti-Nb.

A number of authors have developed classical Nb potentials based on analytic 
functions~\cite{finnis84, ackland87, rebonato87, johnson89, guellil92, 
baskes92, zhang99, lee01, hu02, dai07, wen08, thibaudeau08}. Analytic
potentials are typically fit to experimental data for a small number of 
properties. These potentials reproduce the fit data with high accuracy, but
they often have limited transferability and can produce inaccurate forces for 
molecular dynamics (MD) simulations. The force-matching method proposed by 
Ercolessi and Adams~\cite{ercolessi94} offers an alternative way to construct
potentials. The functions are parameterized by cubic splines, and the spline 
knots are fit to a large number of forces from density-functional theory 
(DFT)~\cite{hohenberg64, kohn65} calculations and usually experimental data as 
well. Including force data from different thermodynamic conditions improves 
accuracy and transferability for a larger range of simulations.

We use the force-matching method to develop a cubic spline-based embedded-atom 
method (EAM) potential~\cite{daw83, daw84} for Nb. The potential is fit to a 
database of DFT forces, energies, and stresses from {\it ab initio} molecular 
dynamics (MD) simulations. We do not fit to any experimental data, since it may
be inconsistent with the DFT information. Instead, we use experimental data and
DFT results to test the accuracy of the potential. Section II discusses our DFT
database calculations and potential optimization process. We utilize the 
force-matching program {\sc potfit}~\cite{brommer06, brommer07} to optimize the
spline knots to the DFT database. Section III presents EAM calculations of 
structural and elastic properties, defects, and thermal behavior. We compare 
the results to DFT calculations and experimental data. These calculations 
demonstrate the potential's ability to describe properties related to the 
fitting data, as well as transferability to behavior beyond the fitting range.

\section{\label{sec:opt} Optimization of the embedded-atom method potential}

\subsection{\label{subsec:EAM} Embedded-atom method interatomic potentials}

EAM potentials~\cite{daw83, daw84} overcome limitations associated with simple 
pairwise interatomic potentials in simulations of metallic systems. Pair 
potentials yield a number of incorrect predictions for transition 
metals~\cite{carlsson90}, including bond energies that are independent of the 
local bonding environment, a zero value for the Cauchy pressure 
($C_{12} - C_{44} = 0$), and the equivalence of the cohesive energy with the 
unrelaxed vacancy formation energy. Supplementing the pairwise interaction with
a volume-dependent term removes some of these undesirable 
features~\cite{carlsson90, johnson72}, but the volume is ill-defined near 
defects such as cracks and surfaces. EAM potentials overcome these difficulties
by implicitly including many-body interactions, and requiring the local 
``electronic density'' as input rather than volume.

The EAM formalism is based on ideas from DFT~\cite{hohenberg64, kohn65}. The 
energy required to embed an impurity atom $Z$ in a solid at position 
${\mathbf R}$ is a unique functional $\mathcal{F}_{Z,\mathbf{R}}[n]$ of the 
electronic density $n$ of the solid before the impurity is 
added~\cite{stott80, norskov82}. The embedded-atom method views each atom in 
the solid as an impurity embedded in a host solid made up of the remaining 
atoms. The energy functional is approximated by a potential energy function 
with two terms: (1) a sum of pairwise interactions 
$\phi_{s_i s_j}(|{\mathbf r}_i - {\mathbf r}_j|)$ between atoms $i$ and $j$, 
and (2) a sum of embedding energies $F_{s_i}(n_i)$ for each atom $i$ that 
depend on the {\it local} electronic density $n_i$ the atom sees from its 
surrounding neighbors. This local electronic density is a sum of radially 
symmetric electronic-density functions $\rho_{s_j}(|{\mathbf r}_i - 
{\mathbf r}_j|)$ arising from the atoms $j$ surrounding a given atom $i$,

\begin{equation}\label{eqn:eam_density}
n_i = \sum_{j \neq i} \rho_{s_j} (r_{ij}),
\end{equation}

\noindent 
where $r_{ij} = |{\mathbf r}_i - {\mathbf r}_j|$ is the distance between atoms 
$i$ and $j$. The total potential energy is

\begin{equation}\label{eqn:eam_energy}
E = \sum_{i<j} \phi_{s_i s_j} (r_{ij}) + \sum_i F_{s_i} (n_i),
\end{equation}

\noindent
where the subscripts $s_i$ and $s_j$ indicate that the functions depend on the 
species of the atoms. 

Equations (\ref{eqn:eam_density}) and (\ref{eqn:eam_energy}) are general and 
hold for multicomponent systems. The energy expression simplifies for monatomic
systems,

\begin{equation}\label{eqn:eam_mono_energy}
E = \sum_{i<j} \phi (r_{ij}) + \sum_i F(n_i),
\end{equation}

\noindent
where

\begin{equation}\label{eqn:eam_mono_density}
n_i = \sum_{j \neq i} \rho (r_{ij}).
\end{equation}

\noindent
Thus, for a single component system the three functions $\phi$, $F$, and $\rho$
must be determined (whereas two-component alloys require seven functions).
EAM potentials are implemented in many freely-available MD codes, e.g., 
{\sc imd}~\cite{stadler97, roth00}, {\sc lammps}~\cite{plimpton95}, and 
{\sc ohmms}~\cite{jkim04}. We perform our EAM calculations using all three of 
these codes.

\subsection{\label{subsec:DFT} Database of DFT forces, energies, and 
stresses}

We use the force-matching method of Ercolessi and Adams~\cite{ercolessi94} to 
obtain accurate potentials for molecular dynamics simulations. Force-matched 
potentials are fit to forces from DFT calculations and typically physical 
properties from either DFT calculations or experiment. Here we include only DFT
data in our fitting database to avoid conflicting information. We use the 
force-matching program {\sc potfit}~\cite{brommer06, brommer07} to optimize the
EAM functions to a database of DFT forces, energies per atom, and stresses for
Nb from the configurations listed in Table~\ref{tab:database}. Fitting to DFT 
data from configurations under different temperature and strain conditions 
extends the applicability of the potential to a wide range of simulations.

\begin{table*}
\caption{\label{tab:database} Configurations in the fitting database. The 
``structure'' column lists the crystal structure of each configuration. The 
``primitive'' and ``conventional'' labels indicate if the supercell is based 
on the one-atom primitive cell or the two-atom conventional cell. The liquid 
configuration starts as a bcc lattice and then melts during the {\it ab initio}
MD simulation. The ``$N_{\rm atoms}$'' column lists the number of atoms in each
configuration. The ``$T$'' column lists the temperatures of the {\it ab initio}
MD simulations used to generate the configurations. The ``$V/V_0$'' column 
lists the ratio of the supercell volume to the zero-temperature, equilibrium 
DFT volume of the bcc, fcc, or hcp structure, respectively. For configuration 
13, $V_0$ is the equilibrium volume of the bcc structure. The ``strain'' column
indicates the strain applied to the supercells, where $M$ denotes a 
volume-conserving monoclinic strain and $O$ denotes a volume-conserving 
orthorhombic strain. The ``rms deviation'' column lists the weighted relative 
rms deviations of the EAM force magnitudes from the DFT values. The 
``$\theta_{\rm avg}$'' column lists the weighted average angular deviation of 
the EAM force directions from the DFT force directions. The weighted relative 
rms force-magnitude deviation and weighted average angular deviation of the 
forces for all the configurations are 25\% and $15.1^\circ$, respectively.}
\begin{ruledtabular}
\begin{tabular}{cccccccc}
Configuration & Structure & $N_{\rm atoms}$ & $T \; {\rm(K)}$ & $V/V_0$ & 
Strain & rms deviation (\%) & $\theta_{\rm avg}$ (deg.) \\ \hline
1 & bcc, primitive & 125 & 300 & 0.90 & None & 18 & 11.7 \\
2 & bcc, primitive & 125 & 300 & 1.00 & None & 20 & 12.3 \\
3 & bcc, primitive & 125 & 300 & 1.10 & None & 29 & 18.4 \\
4 & bcc, primitive, vacancy & 124 & 300 & 1.00 & None & 38 & 17.9 \\
5 & bcc, conventional & 128 & 300 & 1.00 & 2\%, $M$ & 20 & 14.3 \\
6 & bcc, conventional & 128 & 300 & 1.00 & 1\%, $M$ & 22 & 16.1 \\
7 & bcc, conventional & 128 & 300 & 1.00 & $-1$\%, $M$ & 21 & 14.0 \\
8 & bcc, conventional & 128 & 300 & 1.00 & $-2$\%, $M$ & 23 & 15.0 \\
9 & bcc, conventional & 128 & 300 & 1.00 & 2\%, $O$ & 23 & 16.0 \\
10 & bcc, conventional & 128 & 300 & 1.00 & $-2$\%, $O$ & 21 & 15.9 \\
11 & bcc, primitive & 125 & 1,200 & 1.00 & None & 21 & 13.2 \\
12 & bcc, primitive & 125 & 2,200 & 1.00 & None & 19 & 12.4 \\
13 & liquid, primitive & 125 & 5,000 & 1.00 & None & 21 & 11.9 \\
14 & fcc, primitive & 125 & 300 & 1.00 & None & 36 & 29.0 \\
15 & hcp, conventional & 128 & 300 & 1.00 & None & 51 & 37.4 \\
\end{tabular}
\end{ruledtabular}
\end{table*}

The DFT calculations are performed using the plane-wave program 
{\sc vasp}~\cite{kresse96}. The Perdew-Burke-Ernzerhof (PBE) 
generalized-gradient approximation (GGA) functional~\cite{perdew96} accounts 
for the electronic exchange-correlation energy, and a projector augmented-wave 
(PAW) pseudopotential~\cite{blochl94} generated by Kresse~\cite{kresse99} 
represents the nucleus and core electrons. Along with the five valence states, 
the 4$s$ and 4$p$ semicore states are treated explicitly to accurately 
describe interactions at small interatomic separations. The elastic constants 
also agree better with experiment when more electronic states are included.

The database is calculated in two steps. First, {\em ab initio} (MD) 
simulations generate realistic atomic trajectories for various temperature and 
strain conditions. The simulation supercells contain 124-128 atoms, depending 
on the structure. These calculations use a relatively low convergence criteria 
to reduce the computational burden. A single $k$ point is used, and the 
plane-wave cutoff energy is set to the default value of $219.927 
\; {\rm eV}$ from the {\sc vasp} pseudopotential file. Order-one 
Methfessel-Paxton smearing~\cite{methfessel89} is used with a smearing width of
0.10 eV. The MD simulations run for 400 steps with a 3 fs time-step.

Second, well-converged calculations determine the forces, energies per atom, 
and stresses for the atomic configurations resulting from the final step of the
MD simulations. We use $\Gamma$-centered $k$-point meshes with $30 \times 30 
\times 30$ points per primitive cell and increase the plane-wave cutoff energy
to 550 eV. The value of the smearing parameter is unchanged. The energies are 
converged to less than 1 meV/atom. The fitting database contains 1,895 forces 
(5,685 force components), 15 energies per atom, and 90 stress tensor components
from these calculations. Table~\ref{tab:database} lists the configurations in 
the database, along with the weighted relative rms deviations of the EAM force 
magnitudes from the DFT values and the weighted average angular deviations of 
the EAM force directions from the DFT force directions (these errors are 
discussed in Sec.~\ref{subsec:fit_error}).

\subsection{\label{subsec:opt} Optimization of EAM functions to DFT data}

Generating accurate potentials using the force-matching method is an 
optimization problem in a high-dimensional space. The EAM functions are 
parameterized by cubic splines, and the program {\sc potfit}~\cite{brommer06, 
brommer07} optimizes the spline knots using a combination of simulated 
annealing and conjugate-gradientlike algorithms. Our potential construction
procedure proceeds iteratively. We generate a database of DFT calculations 
and choose an initial set of spline knots. We also specify the cutoffs for the 
functions and the fitting weights for the values in the database. Then the
optimization algorithms in {\sc potfit} adjust the spline knots to minimize the
weighted error between the database values and the corresponding values 
produced by the EAM potential. If the fitting errors are too large and the
potential fails to produce satisfactory results for physical properties, we 
add or remove configurations from the database, change the fitting weights and
cutoffs, and/or modify the number and initial values of the spline knots, and 
refit the potential. This optimization and testing process is repeated 
until we obtain accurate potentials.

In {\sc potfit}, the fitting error is defined through a least-squares target 
function formed from the differences between the EAM and DFT values:

\begin{equation}\label{eqn:err}
{\mathcal Z} = {\mathcal Z}_{\rm F} + {\mathcal Z}_{\rm C},
\end{equation}

\noindent
where

\begin{equation}\label{eqn:force_err}
{\mathcal Z}_{\rm F} = \sum_{i = 1}^{N_{\rm A}} \sum_{j = 1}^3 W_i 
\frac{\left(F_{i, x_j}^{\rm EAM} - F_{i, x_j}^{\rm DFT}\right)^2}
{\left(F_{i, x_j}^{\rm DFT}\right)^2 + \epsilon_i},
\end{equation}

\noindent
and

\begin{equation}\label{eqn:global_err}
{\mathcal Z}_{\rm C} = \sum_{i = 1}^{N_{\rm C}} W_i \frac{\left(A_i^{\rm EAM} -
A_i^{\rm DFT}\right)^2}
{\left(A_i^{\rm DFT}\right)^2 + \epsilon_i}.
\end{equation}

\noindent
Equation~(\ref{eqn:force_err}) is the relative deviation of the EAM forces from
the DFT forces, where $N_{\rm A}$ is the number of atoms in the fitting 
database, $F_{i, x_j}$ is the ${x_j}{\rm th}$ component of the force on atom 
$i$, $W_i$ is the weight associated with each force, and $\epsilon_i$ is a 
small number that prevents overweighting of very small, inaccurate forces. 
Equation~(\ref{eqn:global_err}) is the relative deviation between the EAM 
energies per atom and stresses and the DFT values, where $N_{\rm C}$ is the 
number of energies per atom and stress tensor components, $A_i$ is an energy or
stress value, $W_i$ is the associated weight, and $\epsilon_i$ is a small 
number that prevents overweighting of numerically small data. The optimal 
spline knots minimize ${\mathcal Z}$. See Brommer and 
G{\"{a}}hler~\cite{brommer06, brommer07} for {\sc potfit} details.

We determine if there are more parameters in the potential than the fitting
database can support, i.e., over-fitting, by calculating the errors for a 
testing database of DFT forces, energies per atom, and stresses for bcc, fcc, 
and hcp configurations not included in the fitting database. If the errors for 
the testing database are much larger than the errors for the fitting database, 
there are likely too many parameters specifying the EAM 
functions~\cite{robertson93, mishin99, brommer07}. We also test the optimized 
potential's ability to predict physical properties (see Sec.~\ref{sec:RandD}). 
If the potential fails to adequately describe the databases and desired 
properties, we add or remove configurations from the databases, modify the 
fitting weights and cutoffs, and/or change the number and initial values of the
spline knots, and the optimization and testing process repeats. 

Typical of simulated annealing methods, several hundred iterations of this 
procedure were required to find a small number of reasonable potentials. For 
the fitting database listed in Table~\ref{tab:database}, we find that 
potentials with 15-20 spline knots for $\phi$ and $\rho$ and 5-10 spline knots 
for $F$, and a cutoff radius of 4.75 {\AA} for $\phi$ and $\rho$ produce the 
most physically reasonable results, while giving similar force-matching errors 
of 20-30\% for the fitting and testing databases. The cutoff radius includes 
first, second, and third nearest-neighbor interactions in bcc Nb. The cutoffs 
for $F$ are updated automatically by {\sc potfit} as $\rho$ changes during 
optimization.

\subsection{\label{subsec:best_pot} Optimized potential}

\begin{figure*}
\includegraphics[width=0.95\textwidth]{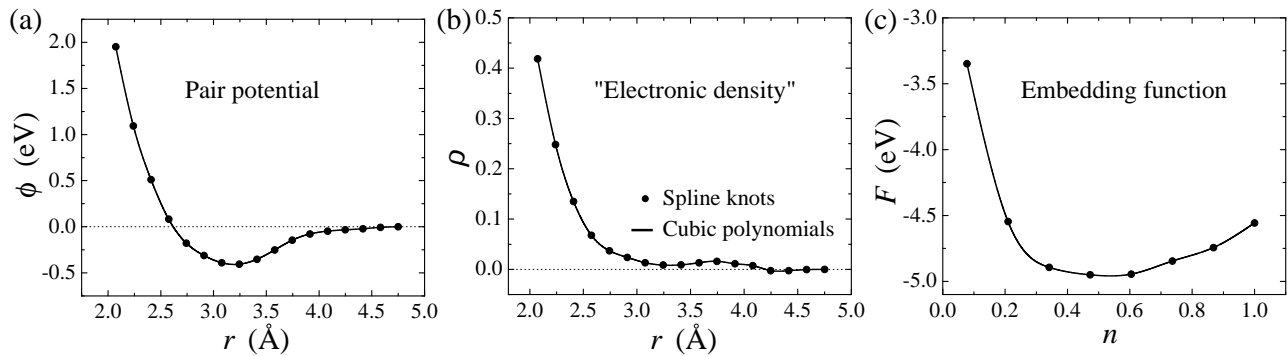}
\caption{\label{fig:splines} The three cubic splines of the EAM potential. The 
points are optimized spline knots and the solid lines are cubic polynomials 
that interpolate between the knots. (a) The pair potential $\phi$ and (b) the 
``electronic density'' $\rho$ are functions of the distance $r$ between pairs 
of atoms. Both of these functions have seventeen optimized spline knots and a 
cutoff radius of 4.75 \AA, which includes first-, second-, and 
third-nearest-neighbor interactions in bcc Nb. (c) The embedding function $F$ 
depends on the local electronic density $n$. Eight optimized spline knots 
parameterize $F$.}
\end{figure*}

Figure~\ref{fig:splines} shows the optimized cubic splines of the best Nb 
potential that we found. The pair potential $\phi$ and electronic density
$\rho$ are parameterized using seventeen equally spaced spline knots while 
eight spline knots are used for the embedding function $F$. The outer cutoff 
distance for $\phi$ and $\rho$ is 4.75 \AA. The shortest interatomic distance 
in the fitting database, which is 2.073 \AA, determines the inner cutoff 
distance. The inner and outer cutoffs of $F$ are 0.0775 and 1.000, 
respectively.

The function $\phi$ has the expected characteristics for a pair potential. The 
interaction is attractive for large interatomic separations, and highly 
repulsive when atoms approach too closely. The minimum value for $\phi$ occurs 
at $r = 3.197 \; {\rm \AA}$. The electronic density $\rho$ is large for 
small interatomic separations and decreases for $r$-values up to about 3.25 
\AA. Beyond this distance, $\rho$ ripples and decays to zero at 4.75 \AA. The 
zero-temperature equilibrium value of $n$ is 0.263. We find that the 
non-monotonic character of $\rho$ is required for an accurate description of 
Nb. Potentials with smoother $\rho$ functions found during the optimization 
and testing procedure yield poor results for many properties. The embedding
function $F$ has positive curvature over most of its range, but there is a 
small region of negative-curvature around $n = 0.74$. This behavior is not 
ideal but atoms rarely sample $n$-values greater than 0.6 even at large 
temperatures and pressures.

In addition to specifying the spline knots and requiring continuity of the 
first and second derivatives of the functions at the knots, two boundary 
conditions must be applied to each function to determine all the cubic 
polynomial coefficients. The natural boundary condition, i.e., a vanishing 
second derivative, is applied at the inner cutoff radius of $\phi$ and $\rho$, 
and at the inner and outer cutoffs of $F$. The remaining boundary conditions 
are the first derivatives of $\phi$ and $\rho$ are zero at the outer cutoff 
radius. Appendix discusses modifications to $\phi$, $\rho$, and $F$ for small 
and large values of their arguments. These modifications improve the 
performance of the potential at large temperatures and pressures. 
Table~\ref{tab:knots} lists the spline knots and boundary conditions for 
$\phi$, $\rho$, and $F$. The potential functions are available in tabulated
form upon request.

\begin{table*}
\caption{\label{tab:knots} The cubic spline knots and boundary conditions. 
Spline knots 1-17 for $\phi$ and $\rho$, and spline knots 1-8 for $F$ are 
optimized by {\sc potfit}. The adjusted values of knot 0 for $\phi$, $\rho$, 
and $F$, and knot 9 for $F$ are also listed (see Appendix). The coefficients of
the cubic polynomials that interpolate between the knots are determined by 
requiring continuity of the functions and their first and second derivatives, 
along with the boundary conditions.
}
\begin{ruledtabular}
\begin{tabular}{cccccc}
$i$ & $r_i$ (\AA) & $\phi(r_i)$ (eV) & $\rho(r_i)$ & $n_i$ & $F(n_i)$ (eV) \\ 
\hline
0 & 1.7383750 & 5.644808063640994 & 0.683176019233847 & 
0.000000000000000 & 0.000000000000000 \\
1 & 2.0730000 & 1.952032491449762 & 0.418661384304128 & 
0.077492938439077 & $-3.347285692522362$ \\
2 & 2.2403125 & 1.094035979464646 & 0.248142385672424 & 
0.209279661519209 & $-4.546334492745762$ \\
3 & 2.4076250 & 0.510885854762808 & 0.135151131573890 & 
0.341066384599341 & $-4.893456225397550$ \\
4 & 2.5749375 & 0.082343335887366 & 0.067802030440920 & 
0.472853107679473 & $-4.950236437181159$ \\
5 & 2.7422500 & $-0.177550651790219$ & 0.037078599738033 & 
0.604639830759605 & $-4.944970691786193$ \\
6 & 2.9095625 & $-0.311331931736446$ & 0.023834158891363 & 
0.736426553839736 & $-4.845699482076931$ \\
7 & 3.0768750 & $-0.390004380615947$ & 0.013226669087316 & 
0.868213276919868 & $-4.743717588952880$ \\
8 & 3.2441875 & $-0.405551151985570$ & 0.008594239037838 & 
1.000000000000000 & $-4.556142211118433$ \\
9 & 3.4115000 & $-0.351882201216042$ & 0.009026077313542 & 
1.263573446160264 & 4.828348385154062 \\
10 & 3.5788125 & $-0.251634925091355$ & 0.013228711231271 & & \\
11 & 3.7461250 & $-0.145378019920633$ & 0.016102598867695 & & \\
12 & 3.9134375 & $-0.078119761728408$ & 0.011199412726043 & & \\
13 & 4.0807500 & $-0.047220500113419$ & 0.007407238328861 & & \\
14 & 4.2480625 & $-0.032830828537903$ & $-0.002416625008422$ & & \\
15 & 4.4153750 & $-0.021236531023427$ & $-0.002572474995293$ & & \\
16 & 4.5826875 & $-0.006495370564318$ & $-0.000515878027624$ & & \\
17 & 4.7500000 & 0.000000000000000 & 0.000000000000000 & & \\ \\
$\;$ & $\;$ & Boundary conditions & \\
$\;$ &$\phi''(r_1) = 0$ &$\phi'(r_{17}) = 0$ & \\
$\;$ & $\rho''(r_1) = 0$ & $\rho'(r_{17}) = 0$ & \\
$\;$ & $F''(n_1) = 0$ & $F''(n_8) = 0$ & \\
\end{tabular}
\end{ruledtabular}
\end{table*}

\subsection{\label{subsec:fit_error} Fitting errors}

\begin{figure*}
\includegraphics[width=0.95\textwidth]{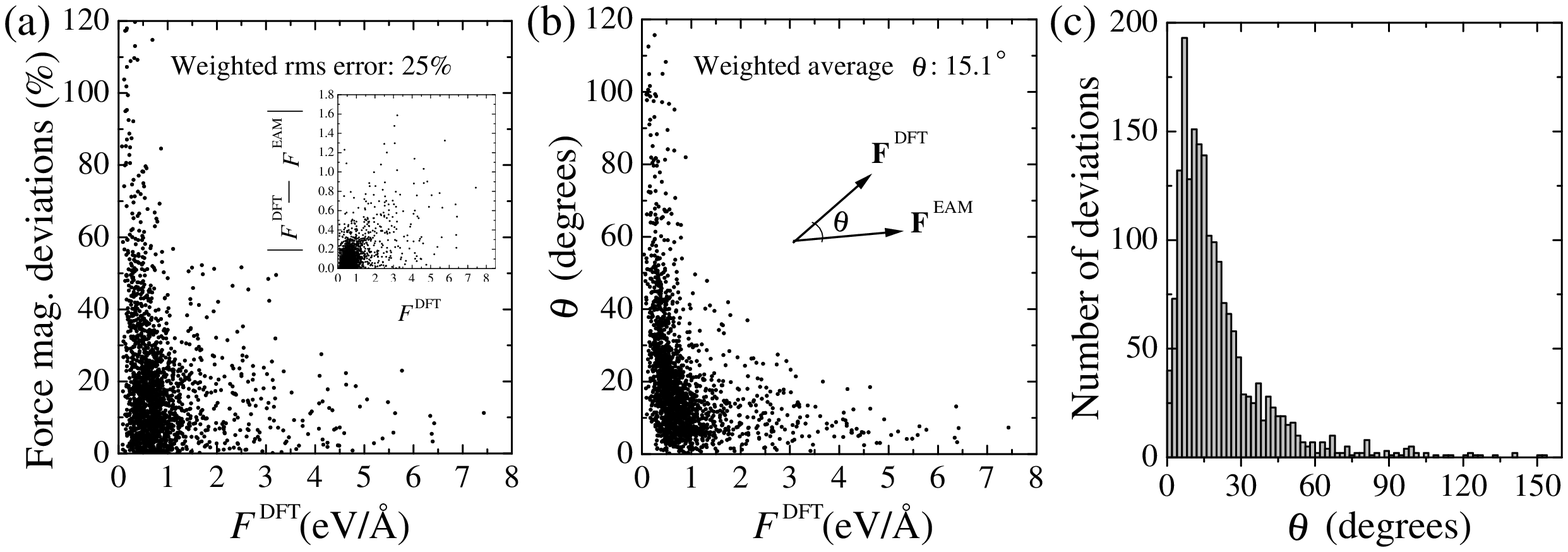}
\caption{\label{fig:errors} Errors between the 1,895 DFT forces in the fitting 
database and the corresponding EAM forces. The relative errors in the forces 
tend to decrease as the force magnitudes increase. (a) The percent relative 
deviation of the EAM force magnitudes from the DFT values, versus the 
magnitudes of the DFT forces. The weighted relative rms deviation of all the 
forces is 25\%. The inset shows the absolute errors in the forces. (b) The 
angular deviations of the EAM force directions from the corresponding DFT force
directions. The weighted average of all the angular deviations is 
$15.1^{\circ}$. (c) Histogram of the angular deviations: 81\% of the deviations
are within the range $0 < \theta < 30^\circ$, 96\% of the deviations are within
the range $0 < \theta < 60^\circ$, and 98\% of the deviations are within the 
range $0 < \theta < 90^\circ$.} \end{figure*}

The fitting database contains DFT forces, energies, and stresses for the
configurations listed in Table~\ref{tab:database}. The EAM potential computes 
the same set of quantities for the fixed atomic positions of each 
configuration and we evaluate the deviations of the EAM values from the 
corresponding DFT values. The errors associated with numerically small DFT data
are typically much greater than the errors from larger data. This is 
illustrated in Figs.~\ref{fig:errors}(a) and~\ref{fig:errors}(b), which show 
the relative force-magnitude deviation and angular deviation of each of the EAM
forces from the DFT database values, versus the DFT force magnitudes. Since 
very small values are inherently inaccurate, we weight the terms in the error 
calculations by the magnitudes of the DFT values. The weighted relative rms 
deviation of the energies per atom, stresses, or force magnitudes is 

\begin{equation}\label{eqn:rms}
{\Delta Q}_{\rm rms} = \sqrt{\sum_{i=1}^{N_{\rm Q}} \omega_i 
\left(\frac{{Q_i}^{\rm EAM}-{Q_i}^{\rm DFT}}{{Q_i}^{\rm DFT}}\right)^2} \times 
100\%,
\end{equation}
 
\noindent
where $Q_i$ is an energy per atom, a stress tensor component, or a force 
magnitude, and $N_{\rm Q}$ is the respective number of such quantities in the 
database. The scaled magnitudes of the DFT data $\omega_i$ weight the terms in
the sum,

\begin{equation}\label{eqn:weights}
\omega_i = \frac{\left|{Q_i}^{\rm DFT}\right|}{\sum_{j = 1}^{N_{\rm Q}} 
\left|{Q_j}^{\rm DFT}\right|}.
\end{equation}

The EAM potential reproduces the energies per atom of the configurations with a
weighted rms deviation of only 0.1\%. The diagonal components of the stress 
tensors are also accurately reproduced with a 6\% weighted rms deviation. In 
contrast, the error for the off-diagonal components of the stress tensors is 
very large. The weighted rms deviation for these quantities is 307\%. The 
off-diagonal values are very small however, even for the strained supercells. 
Increasing the strain yields larger stress values, but the DFT stress-strain 
curves for Nb become non-linear for strains larger than about 2.5\%. 
Table~\ref{tab:database} lists the weighted relative rms deviations of the 
force magnitudes for each configuration in the database. The weighted relative 
rms deviation for all the configurations is 25\%. This force-magnitude error is
lower than the 32\% error of Li {\it et al.}'s force-matched tantalum EAM 
potential~\cite{li03}, and similar to the error of Hennig {\it et al.}'s 
force-matched titanium modified EAM (MEAM) potential~\cite{hennig08}. A direct 
comparison of the errors is difficult, however, due to the different types of 
potentials and/or configurations considered in each work.

We also determine the errors for the directions of the forces. The weighted 
average angular deviation of the EAM force directions from the DFT force 
directions is

\begin{equation}\label{eqn:theta_avg}
\theta_{\rm avg} = \sum_{i = 1}^{N_{\rm atoms}} \omega_i \theta_i,
\end{equation}

\noindent
where $\theta_i$ is the angle between the EAM force on atom $i$ and the DFT 
force on atom $i$. Each angle in the sum is weighted by the corresponding 
scaled DFT force magnitude $\omega_i$. Table~\ref{tab:database} lists the 
weighted average angular deviation of the forces for each database 
configuration. The weighted average angular deviation for all the 
configurations is only $15.1^{\circ}$. The histogram in 
Fig.~\ref{fig:errors}(c) shows that 81\% of the angular deviations are less 
than $30^\circ$, and 98\% of the angular deviations are less than $90^\circ$.

The testing database contains 1,381 forces from nine bcc configurations, one 
fcc configuration, and one hcp configuration. The data is generated for 
temperatures and pressures that lie between and beyond the temperatures and 
pressures in the fitting database. The weighted relative rms deviation of the 
EAM force magnitudes from the DFT values is 27\%, and the weighted average 
angular deviation of the EAM force directions from the DFT force directions is 
16.5$^\circ$. These values are very similar to the fitting database errors,
indicating that the fitting database contains enough data to support the number
of parameters in the potential.


\section{\label{sec:RandD} Results and Discussion}

We assess the quality of the potential by comparing a wide variety of computed 
properties to DFT calculations and experimental data. All the DFT calculations 
use the same method and convergence criteria as the database calculations: PBE 
exchange-correlation functional, PAW pseudopotential with valence states and 
4$s$ and 4$p$ semicore states treated explicitly, $\Gamma$-centered $k$-point 
meshes with $30 \times 30 \times 30$ points per primitive cell, a plane-wave 
cutoff energy of 550 eV, and order-one Methfessel-Paxton smearing with a 
smearing width of 0.10 eV. We calculate two classes of properties: (1) 
properties such as elastic constants which are directly related to 
configurations included in the fitting database, and (2) properties such as 
surface energies which are not related to configurations included in the 
fitting database. The second class serves to test for over-fitting and 
transferability. The potential performs well in nearly all situations we have 
tested.

\subsection{\label{subsec:SandE} Structural and elastic properties}

The potential's first test is reproducing the cohesive energy, lattice 
parameter, and elastic properties of bcc Nb. We also determine the energetic 
stability of the bcc lattice with respect to several other crystal structures. 
Table~\ref{tab:basic} compares the EAM results to our DFT calculations and 
experimental data. The cohesive energy, lattice parameter, and bulk modulus are
determined by calculating the energy of bcc Nb for the volume range $0.90 V_0 <
V < 1.10 V_0$, where $V_0$ is the equilibrium volume, and fitting the 
third-order Birch-Murnaghan equation of state~\cite{birch78, mehl93, BMnote} to
the results. DFT produces a cohesive energy 6\% lower than the experimental 
value~\cite{kittel96}. The EAM cohesive energy is slightly different than the 
DFT value, since the DFT energies per atom of several structures under 
different thermodynamic conditions are used to construct the potential rather 
than the zero-temperature energy per atom. Both DFT and the EAM potential 
reproduce the lattice parameter measured at 4.2 K~\cite{roberge75} with an 
error of less than 1\%. The DFT and EAM bulk modulus values closely match the 
experimental result~\cite{simmons71}, each with an error of less than 1\%. The 
bulk modulus for cubic crystals is related to the elastic constants $C_{11}$ 
and $C_{12}$ via $B = (C_{11} + 2 C_{12})/3$.

\begin{table}
\caption{\label{tab:basic} The EAM values for the cohesive energy, lattice 
parameter, bulk modulus, and elastic constants of bcc Nb are compared to DFT 
and experiment. The experimental lattice parameter and elastic constants were 
measured at 4.2 K. The EAM values for the energies and lattice parameters of 
the fcc, hcp, $\beta$-W, $\beta$-Ta, and $\omega$-Ti structures are compared to
DFT results. The energies are relative to the energy of the bcc structure.}
\begin{ruledtabular}
\begin{tabular}{cccc}
$\;$ & EAM\footnotemark[1] & GGA-PBE\footnotemark[1] & Experiment \\ \hline
$E_{\rm coh}$ (eV/atom) & 7.09 & 7.10 & 7.57\footnotemark[2] \\
$a$ (\AA) & 3.308 & 3.309 & 3.303\footnotemark[3] \\
$B$ (GPa) & 172 & 172 & 173\footnotemark[4] \\
$C_{11}$ (GPa) & 244 & 251 & 253\footnotemark[4] \\
$C_{12}$ (GPa) & 136 & 133 & 133\footnotemark[4] \\
$C_{44}$ (GPa) & 32 & 22 & 31\footnotemark[4] \\
$\Delta E_{\rm fcc-bcc}$ (meV/atom) & 187 & 324 & ... \\
$a_{\rm fcc}$ (\AA) & 4.157 & 4.217 & ... \\
$\Delta E_{\rm hcp-bcc}$ (meV/atom) & 187 & 297 & ... \\
$a_{\rm hcp}$ (\AA) & 2.940 & 2.867 & ...\\
$c_{\rm hcp}$ (\AA) & 4.800 & 5.238 & ...\\
$\Delta E_{\beta{\rm W-bcc}}$ (meV/atom) & 77 & 104 & ... \\
$a_{\beta{\rm W}}$ (\AA) & 5.280 & 5.296 & ... \\
$\Delta E_{\beta{\rm Ta-bcc}}$ (meV/atom) & 105 & 83 & ... \\
$a_{\beta{\rm Ta}}$ (\AA) & 10.200 & 10.184 & ...\\
$c_{\beta{\rm Ta}}$ (\AA) & 5.313 & 5.371 & ...\\
$\Delta E_{\omega{\rm Ti-bcc}}$ (meV/atom) & 167 & 201 & ... \\
$a_{\omega{\rm Ti}}$ (\AA) & 4.845 & 4.887 & ...\\
$c_{\omega{\rm Ti}}$ (\AA) & 2.735 & 2.678 & ...\\
\end{tabular}
\footnotetext[1]{This work.}
\footnotetext[2]{Experimental data from Kittel~\cite{kittel96}.}
\footnotetext[3]{Experimental data from Roberge~\cite{roberge75}.}
\footnotetext[4]{Experimental data from Simmons and Wang~\cite{simmons71}. 
The bulk modulus is obtained from $C_{11}$ and $C_{12}$: $B = (C_{11} + 2 
C_{12}) / 3$.}
\end{ruledtabular}
\end{table}

We compute the elastic constants $C' = (C_{11} - C_{12})/2$ and $C_{44}$ by
straining the bcc crystal and calculating the resulting stress. The slopes of 
the stress versus stain curves yield the elastic constants. We use a 
volume-conserving orthorhombic strain to compute $C'$, and a volume-conserving 
monoclinic strain for $C_{44}$~\cite{mehl93}. We apply a range of strains from 
$-1\%$ to $+1\%$ in each case. $B$ and $C'$ determine $C_{11}$ and $C_{12}$. 
The errors of the EAM elastic constants compared to experiment~\cite{simmons71}
are 4\%, 2\%, and 3\% for $C_{11}$, $C_{12}$, and $C_{44}$, respectively. The 
measured values are from single crystals at 4.2 K. In principle, the EAM value
for $C_{44}$ should closely match the DFT value since no experimental data is
used to fit the potential. The {\sc potfit} program fits to DFT stresses rather
than the elastic constants, and the off-diagonal stress tensor component 
$\sigma_{xy}$ determines $C_{44}$. The off-diagonal stress tensor components 
in the fitting database are generally much smaller than the diagonal 
components, and it is difficult to fit potentials that yield accurate $C_{44}$ 
values. For example, the largest $\sigma_{xx}$ value in the database is 22.4 
GPa while the largest $\sigma_{xy}$ value is only 0.490 GPa. A large stress 
fitting weight must be used to produce potentials with $C_{44}$ values close to
the DFT and experimental values.

The stability of the bcc crystal structure is demonstrated with respect to the 
fcc and hcp structures. DFT predicts that the energies per atom for fcc and hcp
Nb are 323 meV and 296 meV larger than the bcc value, respectively. The EAM 
potential predicts that the energy per atom for both of these structures is 187
meV larger than the bcc value. Our EAM potential produces the ideal 
close-packed $c/a$ value of 1.633 for the hcp structure, whereas the DFT value 
is $c/a = 1.827$. For $c/a = 1.633$, the fcc and hcp first-nearest-neighbor 
distances are equal, as are the second-nearest-neighbor distances. Therefore, 
third-nearest-neighbor interactions must be included to differentiate between 
fcc and hcp for potentials with no angular dependence. Our EAM potential 
includes first-, second-, and third-nearest-neighbor interactions in the bcc
structure, but only first- and second-nearest-neighbor interactions for the fcc
and hcp structures. This leads to the ideal $c/a$ ratio in the hcp structure, 
and energetic degeneracy of the fcc and hcp structures. Increasing the range of
$\phi$ and $\rho$ and including more fcc and hcp data in the fitting database 
produces values closer to the DFT results, but the bcc elastic constants, 
phonon dispersions, and vacancy formation energy agreed poorly with DFT and
experiment.

The bcc metals tungsten (W) and tantalum (Ta) have metastable $\beta$ phases
based on structures with eight atoms per unit cell and thirty atoms per unit 
cell, respectively. The $\beta$-W structure has $Pm \bar3 n$ symmetry, and the 
$\beta$-Ta structure has $P4_2/mnm$ symmetry. Titanium (Ti) transforms from hcp
to bcc at 1,155 K and ambient pressure, and has a high-pressure $\omega$-phase 
based on a three-atom unit cell with $P6/mmm$ symmetry. No data for the 
$\beta$-W, $\beta$-Ta, and $\omega$-Ti structures is included in the fitting 
database. The energies of these structures are higher than the bcc energy. The 
lattice parameters of all the structures are reproduced reasonably well with 
the largest error for the hcp $c$ value. The energetic ordering of the 
structures is different in EAM and DFT but bcc is most stable in both cases. 
We also find that in finite-temperature MD simulations, the EAM potential
stabilizes the bcc structure to the melting point for pressures below 125 GPa 
(see Sec.~\ref{subsec:lattice}).

\subsection{\label{subsec:point} Point defects}

Vacancy motion is the predominant mechanism for solid-state diffusion, and 
the presence of vacancies influences many material processes including 
dislocation motion and creep. We use our EAM potential to calculate the 
single-vacancy formation energy $E_{\rm vac}^{\rm f}$ and migration energy 
$E_{\rm vac}^{\rm m}$, and the activation energy for vacancy diffusion 
$Q_{\rm vac} = E_{\rm vac}^{\rm f} + E_{\rm vac}^{\rm m}$. The simulation 
supercells contain 8,191 atoms. We determine the vacancy migration energy with 
the nudged elastic band method~\cite{mills94, mills95, jonsson98} using seven 
image configurations between the initial and final configurations. The 
migration path is along the $\langle 111\rangle$ direction. We also compute the
vacancy energies using DFT for supercells with 249 atoms. In all our 
calculations, the atoms are relaxed using the conjugate-gradient 
method~\cite{polak69, press07}.

Table~\ref{tab:vacancy} compares our EAM vacancy energies to our DFT results
and other published EAM~\cite{guellil92, hu02}, Finnis-Sinclair 
(F-S)~\cite{harder86}, and MEAM~\cite{lee01} calculations. Most of the results 
are consistent with the experimental data~\cite{landolt91, ablitzer77, 
einziger78, bussmann81, siegel82}. Our EAM potential produces the largest 
formation energy. The {\sc potfit} program fits to the DFT energy per atom of 
each configuration in the database, instead of fitting to defect energies. The 
DFT energy-difference per atom between an ideal crystal and a crystal with a 
single vacancy is about 10 meV. This is close to the accuracy with which the 
EAM potential reproduces the energies in the fitting database and a large 
energy fitting weight must be used to achieve reasonable results.

\begin{table*}
\caption{\label{tab:vacancy} Single-vacancy formation, migration, and diffusion
activation energies. The activation energy is the sum of the formation and 
migration energies:  $Q_{\rm vac} = E_{\rm vac}^{\rm f} + E_{\rm vac}^{\rm m}$.
Energies are reported in eV.}
\begin{ruledtabular}
\begin{tabular}{ccccccccccc} 
$\;$ & EAM\footnotemark[1] & GGA-PBE\footnotemark[1] & Experiment & 
EAM\footnotemark[2] & EAM\footnotemark[3] & F-S\footnotemark[4] 
& MEAM\footnotemark[5] & MEAM\footnotemark[6] \\ \hline
$E_{\rm vac}^{\rm f}$ & 3.10 & 2.72 & 2.6-3.1\footnotemark[7] & 2.88 & 2.76 & 
2.48 & 2.75 & 2.75 \\
$E_{\rm vac}^{\rm m}$ & 0.77 & 0.55 & 0.6-1.6\footnotemark[7] & 0.97 & 0.64 & 
0.91 & 0.54 & 0.57 \\
$Q_{\rm vac}$ & 3.87 & 3.27 & 3.6-4.1\footnotemark[8] & 3.85 & 3.40 & 3.39 & 
3.29 & 3.32 \\
\end{tabular}
\footnotetext[1]{This work.}
\footnotetext[2]{EAM results of Guellil and Adams~\cite{guellil92}.}
\footnotetext[3]{EAM results of Hu {\it et al.}~\cite{hu02}.}
\footnotetext[4]{F-S results of Harder and Bacon~\cite{harder86}.}
\footnotetext[5]{MEAM results of Zhang {\it et al.}~\cite{zhang99}.}
\footnotetext[6]{MEAM results of Lee {\it et al.}~\cite{lee01}.}
\footnotetext[7]{Experimental data from 
Landolt-B{\"{o}}rnstein~\cite{landolt91}.}
\footnotetext[8]{Experimental data from~\cite{ablitzer77, einziger78, 
bussmann81, siegel82, landolt91}.}
\end{ruledtabular}
\end{table*}

\begin{figure}
\includegraphics[width=0.40\textwidth]{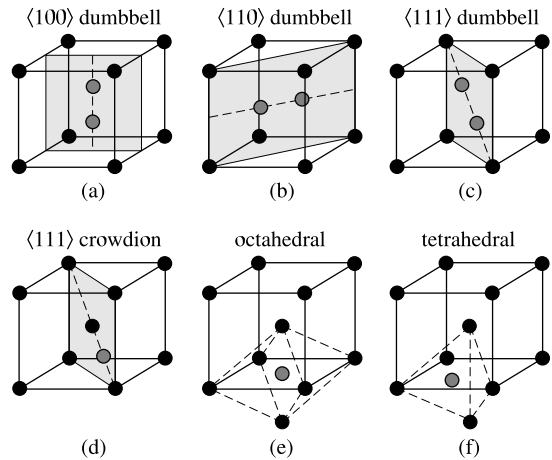}
\caption{\label{fig:interstitials} Schematic illustration of the (a) $\langle
100 \rangle$ dumbbell, (b) $\langle 110 \rangle$ dumbbell, (c) $\langle 111
\rangle$ dumbbell, (d) $\langle 111 \rangle$ crowdion, (e) octahedral, and (f) 
tetrahedral interstitials.}
\end{figure}

In the absence of strong irradiation, the equilibrium concentration of 
self-interstitial atoms in metals is much smaller than the concentration of 
vacancies. Accordingly, no data from configurations with interstitials is 
included in the fitting database. Instead, interstitial formation energy 
calculations can test the transferability of the EAM potential. We determine 
the formation energies of six self-interstitial configurations: the $\langle 
100 \rangle$ dumbbell, $\langle 110 \rangle$ dumbbell, $\langle 111 \rangle$ 
dumbbell, $\langle 111 \rangle$ crowdion, octahedral, and tetrahedral 
interstitials. Figure~\ref{fig:interstitials} shows the geometry of these 
defects. The EAM simulation supercells contain 31,251 atoms which are relaxed 
using the conjugate-gradient method. Since no experimental data is available, 
we also compute the formation energies with DFT. The DFT supercells contain 251
atoms which are relaxed with the conjugate-gradient method. 

Table~\ref{tab:interstitials} lists our EAM results, along with our DFT
values and other published EAM~\cite{hu02}, F-S~\cite{rebonato87, ackland87, 
harder88}, and MEAM~\cite{lee01} results. Our EAM potential yields 
self-interstitial formation energies in the range 3.83-4.50 eV and our DFT 
calculations give formation energies in the range 3.95-4.89 eV. DFT predicts 
that the $\langle 111 \rangle$ dumbbell has the lowest energy while our EAM 
potential predicts the $\langle 110 \rangle$ dumbbell to be the lowest. The EAM
results of Hu {\it et al.}~\cite{hu02} and the F-S results of Ackland and 
Thetford~\cite{ackland87}, Rebonato {\it et al.}~\cite{rebonato87}, and Harder 
and Bacon~\cite{harder88} also place the formation energy of the $\langle 110 
\rangle$ dumbbell lowest. Since our EAM results are not consistent with DFT, 
the potential may not be well suited for radiation damage studies.

\begin{table*}
\caption{\label{tab:interstitials} Formation energies for the $\langle 100 
\rangle$ dumbbell, $\langle 110 \rangle$ dumbbell, $\langle 111 \rangle$ 
dumbbell, $\langle 111 \rangle$ crowdion, octahedral, and tetrahedral 
interstitials in eV. The lowest formation energy for each calculation is 
underlined. No DFT interstitial data is used in constructing our EAM 
potential.}
\begin{ruledtabular}
\begin{tabular}{cccccccccc} 
$\;$ & EAM\footnotemark[1] & GGA-PBE\footnotemark[1] & EAM\footnotemark[2] & 
F-S\footnotemark[3] & F-S\footnotemark[4] & F-S\footnotemark[5] & 
MEAM\footnotemark[6] \\ \hline
$E_{100}^{\rm f}$ & 4.50 & 4.76 & 4.44 & 4.13 & 4.821 & 4.85 & ... \\
$E_{110}^{\rm f}$ & \underline{3.83} & 4.31 & \underline{4.39} & 
\underline{3.99} & \underline{4.485} & \underline{4.54} & 2.56 \\
$E_{111}^{\rm f}$ & 4.09 & \underline{3.95} & 4.74 & ... & 4.795 & 4.88 & ...\\
$E_{\rm crd}^{\rm f}$ & 4.02 & 3.99 & 4.93 & 4.10 & 4.857 & 4.95 & ... \\
$E_{\rm oct}^{\rm f}$ & 4.36 & 4.89 & 4.43 & 4.23 & ... & 4.91 & ... \\
$E_{\rm tet}^{\rm f}$ & 4.37 & 4.56 & 4.73 & 4.26 & ... & 4.95 & ... \\
\end{tabular}
\footnotetext[1]{This work.}
\footnotetext[2]{EAM results of Hu {\it et al.}~\cite{hu02}.}
\footnotetext[3]{F-S results of Rebonato {\it et al.}~\cite{rebonato87}.}
\footnotetext[4]{F-S results of Ackland and Thetford~\cite{ackland87}.}
\footnotetext[5]{F-S results of Harder and Bacon~\cite{harder88}.}
\footnotetext[6]{MEAM result of Lee {\it et al.}~\cite{lee01}.}
\end{ruledtabular}
\end{table*}

\subsection{\label{subsec:lattice} Phonon dispersion, thermal expansion, and
pressure-volume relation}

The next group of properties relate to lattice vibrations and the thermodynamic
behavior of the potential. The calculations demonstrate the applicability of 
the potential over a large range of temperatures and pressures. First, we use 
the program {\sc phon}~\cite{alfe09} to compute the phonon spectra along 
high-symmetry directions in the Brillouin zone. The program employs the 
small-displacement method, in which atoms are moved a small distance from their
equilibrium lattice sites. The dynamical matrix obtained from the resulting 
forces on the atoms yields the phonon dispersions.

\begin{figure*}
\includegraphics[width=0.75\textwidth]{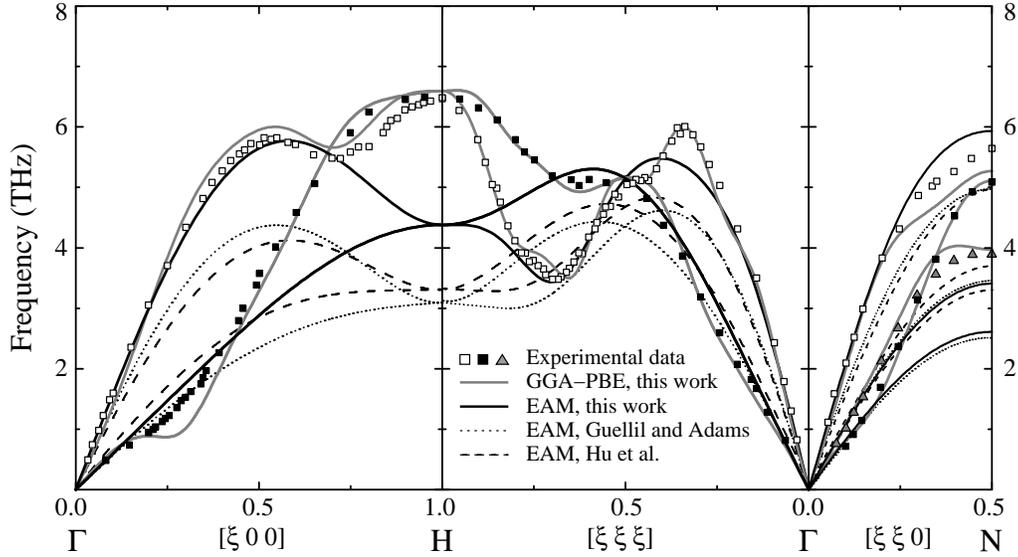}
\caption{\label{fig:phonons} Phonon dispersion curves along high-symmetry 
directions. Our EAM results agree well with experiment~\cite{powell68} for 
small wave vectors, but the classical EAM potential is unable to fully capture 
the spectrum. However, the potential improves upon the EAM results of Guellil 
and Adams~\cite{guellil92} and Hu {\it et al.}~\cite{hu02}. The DFT phonon 
results closely match the experimental data over much of the Brillouin zone. 
The transverse modes in the $[\xi 0 0]$ direction show a plateau around
$\xi = 0.25$ which is an artifact of the interpolation scheme used to
generate the curves. The DFT phonon results were provided by Hennig.}
\end{figure*}

Figure~\ref{fig:phonons} compares the computed phonon spectra along the 
$[\xi 0 0]$, $[\xi \xi \xi]$, and $[\xi \xi 0]$ directions in reciprocal space 
to experimental data~\cite{powell68}, our DFT calculations, and other published
EAM results~\cite{guellil92, hu02}. The DFT calculations are carried out for up
to 512 atoms ($8 \times 8 \times 8$ supercells). The DFT results closely match 
the experimental data over much of the Brillouin zone, but the transverse modes
in the $[\xi 0 0]$ direction show a plateau around $\xi = 0.25$ which is not 
present in the experimental data. This discrepancy is not physical but rather 
is an artifact of the interpolation scheme used in generating the curves. The 
phonon frequencies are computed exactly at only a small number of points in the
Brillouin zone and {\sc phon} interpolates between these exact values to 
generate smooth curves. More exact points, i.e., even larger supercells, are 
required to remove this discrepancy. Our EAM potential accurately describes the
experimental phonon frequencies for small wave-vectors, but is unable to 
reproduce some of the features in the spectrum. This results in poor agreement 
at the zone boundaries H and N but our EAM results match experiment more 
closely than the EAM results of Guellil and Adams~\cite{guellil92} and Hu {\it 
et al.}~\cite{hu02}.

\begin{figure}
\includegraphics[width=0.40\textwidth]{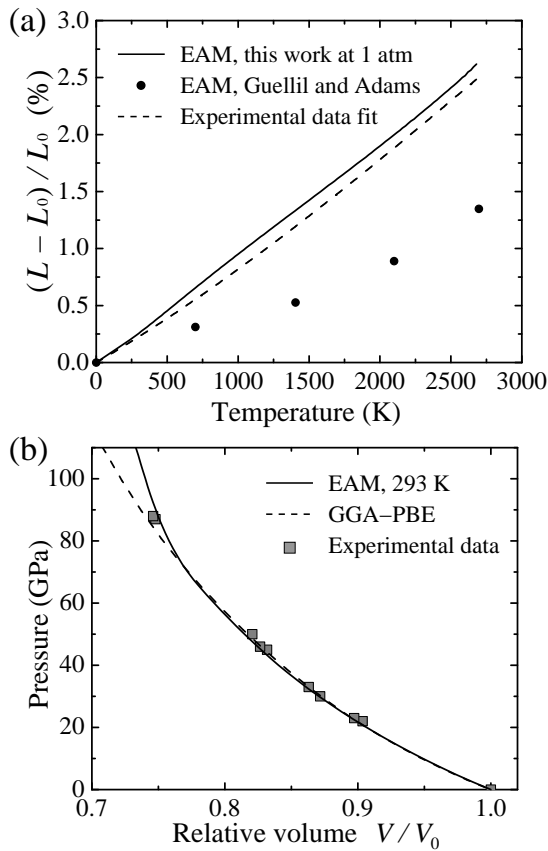}
\caption{\label{fig:latticedynamics} (a) Thermal expansion curve. The thermal 
expansion of our EAM potential agrees closely with 
experiment~\cite{touloukian75} from 0 K to $T_{\rm melt}^{\rm exp} = 2,742 \; 
{\rm K}$. Our EAM potential slightly overestimates the thermal expansion, while
the EAM potential of Guellil and Adams~\cite{guellil92} underestimates it. (b) 
Pressure versus volume curve. The experimental data are from shock 
experiments~\cite{kinslow70}. For pressures to 75 GPa, our EAM potential agrees
well with experimental data and our DFT calculations.}
\end{figure}

Figure~\ref{fig:latticedynamics}(a) shows the thermal expansion of the EAM
potential from 0 K to the experimental melting temperature, 
$T_{\rm melt}^{\rm exp} = 2,742 \; {\rm K}$. Constant-$NPT$ MD simulations of 
8,192 atoms at $P = 1$ atm yield the thermal expansion curve. We determine the
equilibrium lattice constant for 138 temperatures in the range $0 < T < 2,742 
\; {\rm K}$. Each MD simulation runs for 500,000 steps with a 1 fs time step, 
and we determine the lattice constant for each temperature by averaging over 
the last 5,000 simulation steps. We compare the results to experimental 
data~\cite{touloukian75} and the EAM results of Guellil and 
Adams~\cite{guellil92}. Our EAM result lies just above the experimental curve 
while the Guellil and Adams potential underestimates the expansion. Our fitting
database contains data for bcc Nb only at (i) 300 K and $-13$ GPa to 23 GPa, 
(ii) 1,200 K and 2 GPa, and (iii) 2,200 K and 7 GPa, so our results indicate 
the potential accurately interpolates to temperatures and pressures not 
included in the fit.

Figure~\ref{fig:latticedynamics}(b) shows the pressure variation in the EAM 
potential versus the relative volume $V/V_0$, where $V_0$ is the zero-pressure 
volume. Constant-$NPT$ MD simulations of 8,192 atoms at $T = 293$ K yield the 
pressure-volume curve. We determine the equilibrium volume for 50 pressures in 
the range $0 < P < 125 \; {\rm GPa}$. Each MD simulation runs for 500,000 steps
with a 1 fs time step, and we determine the volume for each pressure by 
averaging over the last 5,000 simulation steps. Zero-temperature EAM results 
are nearly identical to the 293 K values. We compare the results to data from 
shock experiments~\cite{kinslow70} and our zero-temperature DFT calculations. 
For pressures to 75 GPa the agreement with DFT and experiment is excellent. The
largest pressure in the fitting database is only 23 GPa from Configuration 1 in
Table~\ref{tab:database}, and the compression of bcc Nb is accurately 
reproduced for more than 50 GPa beyond this pressure. The EAM result deviates 
at larger pressures and at 125 GPa the bcc crystal structure transforms to a 
close-packed lattice. Experiment and DFT do not show a phase transformation. 
Therefore the potential may not be well-suited for shock simulations, but it 
performs very well below 75 GPa. 

\subsection{\label{subsec:surface} Surface properties}

Surface properties test the transferability of our potential to configurations
with low coordination-number since no surface data is used in constructing the
potential. Table~\ref{tab:surfaceE} lists the relaxed surface energies for the 
$\{110\}$, $\{100\}$, and $\{111\}$ surfaces. The EAM calculations use 
slab-geometry supercells with two free surfaces and periodic boundary 
conditions in the directions perpendicular to the surface normals. The 
conjugate-gradient method relaxes 600-layer slabs in directions parallel to the
surface normals while the perpendicular dimensions are fixed. We compare the 
surface energies to our DFT results and published EAM~\cite{guellil92, hu02}, 
F-S~\cite{ackland86}, long-range empirical potential (LREP)~\cite{dai07}, 
MEAM~\cite{baskes92, lee01}, and modified analytical EAM (MAEAM)~\cite{wen08} 
results. Experimental values for energies of individual surfaces are often 
based on simple models, so we evaluate the accuracy of the EAM surface energies
by comparing the results to our DFT calculations.

\begin{table}
\caption{\label{tab:surfaceE} Low-index surface energies of bcc Nb in 
meV/{\rm \AA}$^2$ (J/m$^2$). Our EAM results closely match our DFT values, 
even though no surface data is used to construct the potential.}
\begin{ruledtabular}
\begin{tabular}{cccc} 
$\;$ & $E_{\rm surf}^{\{110\}}$ & $E_{\rm surf}^{\{100\}}$ & 
$E_{\rm surf}^{\{111\}}$
\\ 
\hline
EAM\footnotemark[1] & 127 (2.04) & 147 (2.36) & 154 (2.47) \\
GGA-PBE\footnotemark[1] & 131 (2.10) & 146 (2.34) & 149 (2.39) \\
EAM\footnotemark[2] & 113 (1.81) & 123 (1.97) & ... \\
EAM\footnotemark[3] & 108 (1.73) & 120 (1.93) & ... \\
F-S\footnotemark[4] & 104 (1.67) & 122 (1.96) & ... \\
LREP\footnotemark[5] & 112 (1.79) & 131 (2.10) & 146 (2.34) \\
MEAM\footnotemark[6] & 117 (1.87) & 174 (2.79) & 126 (2.02) \\
MEAM\footnotemark[7] & 155 (2.49) & 169 (2.72) & 182 (2.92) \\
MAEAM\footnotemark[8] & 110 (1.77) & 125 (2.00) & 143 (2.28) \\
\end{tabular}
\footnotetext[1]{This work.}
\footnotetext[2]{EAM results of Guellil and Adams~\cite{guellil92}.}
\footnotetext[3]{EAM results of Hu {\it et al.}~\cite{hu02}.}
\footnotetext[4]{F-S results of Ackland and Finnis~\cite{ackland86}.}
\footnotetext[5]{LREP results of Dai {\it et al.}~\cite{dai07}. Unrelaxed 
surface energies.}
\footnotetext[6]{MEAM results of Baskes~\cite{baskes92}. Unrelaxed surface
energies.}
\footnotetext[7]{MEAM results of Lee {\it et al.}~\cite{lee01}.}
\footnotetext[8]{MAEAM results of Wen and Zhang~\cite{wen08}.}
\end{ruledtabular}
\end{table}

We perform DFT calculation for 24-, 36-, and 48-layer slabs for the $\{100\}$ 
surface, and for 12-, 18-, and 24-layer slabs for the $\{110\}$ and $\{111\}$ 
surfaces. A vacuum region 10 {\AA} thick separates the periodic surface 
images. We relax the slabs in a manner similar to the EAM calculations. We use 
different numbers of layers to study the convergence of the surface energies 
and relaxations with cell size. The energy values for the different numbers of 
layers vary by 1 meV/{\AA}$^2$ or less. When we increase the vacuum layer 
thickness to 15 {\AA} for the largest supercells, the surface energies change 
by less than 0.2 meV/{\AA}$^2$. The errors between our EAM and DFT results for 
the $\{110\}$, $\{100\}$, and $\{111\}$ surfaces are 3.1\%, 0.7\%, and 3.4\%, 
respectively. Both methods predict that $E_{\rm surf}^{\{110\}} 
< E_{\rm surf}^{\{100\}} < E_{\rm surf}^{\{111\}}$. The excellent agreement 
between our EAM and DFT results is surprising, considering the fitting database
does not contain configurations with surfaces. All the potentials give 
reasonable surface energies with respect to the DFT results. The relative rms
deviation of our EAM values from the DFT values is 2.6\%. Dai {\it et al.}'s 
LREP potential~\cite{dai07} has the next lowest relative rms deviation of 
10.3\%, and Lee {\it et al.}'s MEAM potential~\cite{lee01} has the
highest relative rms deviation of 18.9\%. Baskes'~\cite{baskes92} unrelaxed 
MEAM results show a different ordering of the energies than DFT.

Table~\ref{tab:surfaceDelta} lists the percent change in spacing between the 
first and second surface layers relative to the spacing in the bulk. We compare
our EAM results to our DFT calculations, experimental data~\cite{lo98}, and 
published EAM~\cite{guellil92}, F-S~\cite{ackland86}, and MEAM~\cite{lee01} 
results. Our EAM values agree very closely with our DFT calculations. All the 
methods produce contractions of the $\{110\}$, $\{100\}$, and $\{111\}$ surface
layers, except the EAM potential of Guellil and Adams which predicts an 
expansion of the $\{100\}$ layers. Our $\{100\}$ EAM and DFT results also agree
well with experiment. We do not list relaxations for layers deeper beneath the 
surface, since the DFT results oscillate strongly as the number of layers in 
the slab changes.

\begin{table}
\caption{\label{tab:surfaceDelta} Low-index surface relaxations of bcc Nb. The 
values are the relative percent change in the interplanar spacing upon
relaxation. The number of layers in our slabs are in parentheses.}
\begin{ruledtabular}
\begin{tabular}{cccc} 
$\;$ & $\Delta_{12}^{\{110\}} (\%)$  & $\Delta_{12}^{\{100\}} (\%)$ & 
$\Delta_{12}^{\{111\}} (\%)$ \\ 
\hline
EAM\footnotemark[1] & $-5.0 (600)$ & $-13.9 (600)$ & $-27.0 (600)$ \\
GGA-PBE\footnotemark[1] & $-3.9 (12)$ & $-12.4 (24)$ & $-30.7 (12)$ \\
GGA-PBE\footnotemark[1] & $-3.9 (18)$ & $-13.0 (36)$ & $-28.4 (18)$ \\
GGA-PBE\footnotemark[1] & $-4.5 (24)$ & $-12.3 (48)$ & $-27.6 (24)$ \\
Experiment\footnotemark[2] & ... & $-13 \pm 5$ & ... \\
EAM\footnotemark[3] & $-1.6$ & $+0.52$ & ... \\
F-S\footnotemark[4] & $-5.1$ & $-16.0$ & ... \\
MEAM\footnotemark[5] & $-7.3$ & $-12.5$ & $-35.5$ \\
\end{tabular}
\footnotetext[1]{This work.}
\footnotetext[2]{Experimental data from Lo {\it et al.}~\cite{lo98}.}
\footnotetext[3]{EAM results of Guellil and Adams~\cite{guellil92}.}
\footnotetext[4]{F-S results of Ackland and Finnis~\cite{ackland86}.}
\footnotetext[5]{MEAM results of Lee {\it et al.}~\cite{lee01}.}
\end{ruledtabular}
\end{table}

\subsection{\label{subsec:disloc} Stacking faults and dislocations}

The nonplanar core of screw dislocations in bcc transition metals is generally
accepted to be responsible for the complex deformation behavior of these 
materials~\cite{kubin82, christian83, duesbery89, vitek92, seeger95, pichl02, 
duesbery02, groger07}. The cores of $(1/2)\langle 111 \rangle$ screw 
dislocations in bcc metals spread into several planes of the $\langle 111
\rangle$ zone. However, no dissociation of dislocations into well defined 
partial dislocations has been observed, and no metastable stacking faults that 
could participate in such dissociation have been identified. The most widely 
used theoretical approach in searching for possible stacking faults is 
$\gamma$-surface calculations. The $\gamma$ surfaces represent energies of 
generalized stacking faults, formed by displacing two halves of a crystal 
relative to each other along a low-index crystallographic plane~\cite{vitek68},
i.e., the fault plane. As the top half of the crystal moves in the fault plane 
relative to the bottom half, the crystal's ideal stacking order is disrupted. 
The resulting energy increase forms the $\gamma$ surface, which is periodic in 
displacements perpendicular to the fault-plane normal. Minima on 
$\gamma$ surfaces determine possible metastable stacking faults.


We use our EAM potential and DFT to compute sections through the $\{112\}$ and 
$\{110\}$ $ \gamma $ surfaces in the $\langle 111 \rangle$ direction. EAM 
calculations with supercells containing 60,000 atoms determine unrelaxed and 
relaxed $\gamma$-surface energies. The supercell for the $\{112\}$ 
$\gamma$-surface has 3,000 layers and the supercell for the $\{110\}$ 
$\gamma$-surface has 2,000 layers. In each case, the fault-plane divides the 
crystal in half. We calculate the energy as the top half of the crystal is 
displaced relative to the bottom half along $\langle 1 1 1 \rangle$. In the 
relaxed EAM calculations, the atoms are allowed to move only in the direction 
perpendicular to the fault-plane since the stacking faults are unstable. The 
DFT calculations use supercells with 24 layers for the $\{112\}$ fault-plane 
and 12 layers for the $\{110\}$ fault-plane to determine unrelaxed 
$\gamma$-surface energies. 

\begin{figure}
\includegraphics[width=0.40\textwidth]{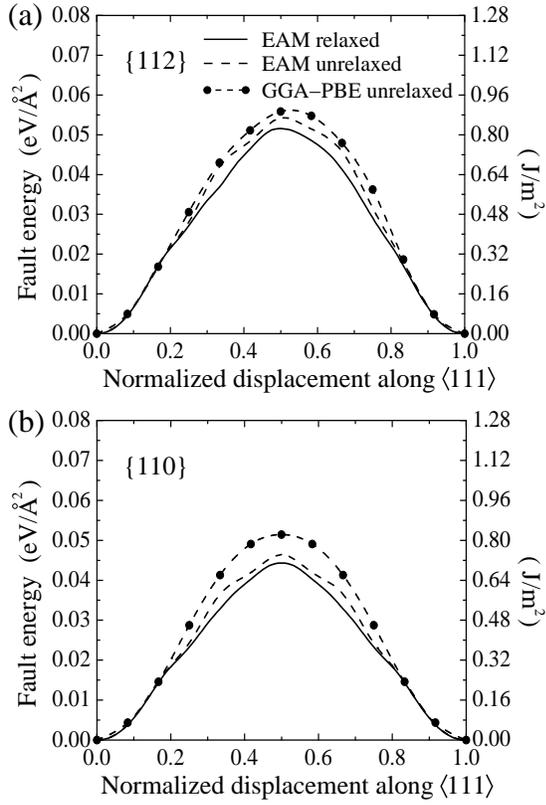}
\caption{\label{fig:gamma} $\gamma$-surface sections in the $\langle 111 
\rangle$ direction. The absence of minima indicate that there are no stable 
stacking faults in the $\{112\}$ and $\{110\}$ planes along this direction. 
The EAM and DFT results agree well, even though no data from configurations 
with stacking faults is used to construct the potential.}
\end{figure}

Figure~\ref{fig:gamma} compares the EAM and DFT $\gamma$-surface sections in 
the $\langle 111 \rangle$ direction for the $\{112\}$ and $\{110\}$ 
fault planes. There are no minima that would indicate the existence of 
metastable stacking faults, which is consistent with $\gamma$-surface 
calculations for many bcc metals~\cite{vitek68, medvedeva96, xu96, duesbery98, 
yang01, frederiksen03}. The overall agreement between our EAM and DFT results 
is very good, but the relaxed (and unrelaxed) EAM results show shoulders near 
$b/6$ and $5b/6$ which are absent in the DFT curves, where $b$ is the magnitude
of the $(1/2)\langle 111 \rangle$ screw dislocation Burgers vector ${\mathbf b}
= (a/2)\langle 111 \rangle$. Relaxed $\gamma$ surfaces from F-S 
calculations~\cite{duesbery98} show similar shoulders for the group VIB element
Mo, but not for the group VB element Ta (Nb is also a group VB element).

\begin{figure}
\includegraphics[width=0.40\textwidth]{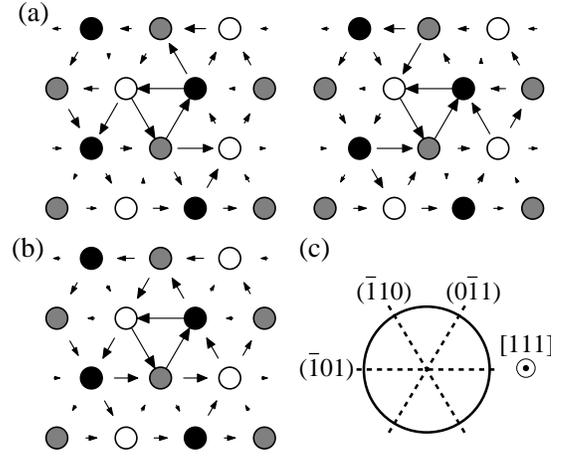}
\caption{\label{fig:cores} Differential-displacement maps of the (a) degenerate
and (b) nondegenerate (or symmetric) core-structures found for $(1/2)[111]$ 
screw dislocations in bcc metals. For the degenerate core, the structures on 
the left and right have the same energy. (c) In all cases, the core spreads 
into three $\{110\}$ planes of the $[111]$ zone.}
\end{figure}

Figure~\ref{fig:cores} shows the two types of $(1/2) \langle 111 \rangle$ screw
dislocation core structures found in calculations for bcc metals. 
Figure~\ref{fig:cores} (a) shows the degenerate core, so named because the 
configurations on the left and right have the same energy. 
Figure~\ref{fig:cores} (b) shows the nondegenerate (or symmetric) core. Both
types of cores spread into three $\{110\}$ planes of the $[111]$ zone. The 
results are presented using differential-displacement maps introduced by Vitek 
{\it et al.}~\cite{vitek70}. The atoms are projected onto the $(111)$ plane, 
and the arrows represent relative atomic displacements in the $[111]$
direction. The lengths of the arrows are scaled such that an arrow connects two
atoms if its length is $b/3$. The shadings of the atoms indicate that there are
three repeating layers of atoms in the $[111]$ direction in an ideal crystal 
(white is the bottom layer and black is the top layer).

\begin{figure}
\includegraphics[width=0.40\textwidth]{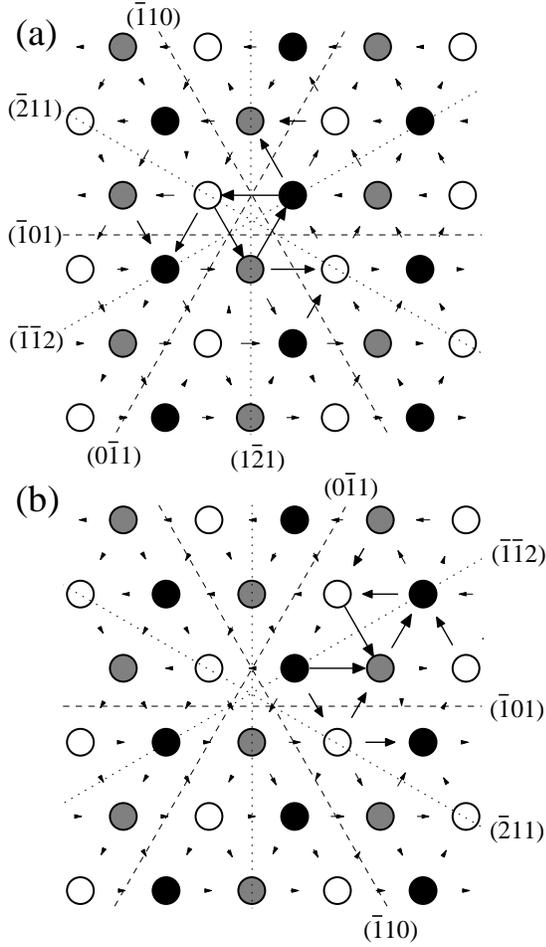}
\caption{\label{fig:screw} Differential-displacement maps for the $(1/2)[111]$
screw dislocation core structure produced by our EAM potential. (a) The 
degenerate core structure results when the atoms are relaxed. (b) Shear stress 
applied along $[111]$ produces a net displacement of the dislocation in the 
$(\bar1 \bar1 2)$ plane. The figure shows the core-structure when the shear
stress is just above the critical-resolved shear stress for dislocation 
motion.} 
\end{figure}

Figure~\ref{fig:screw}(a) shows that our EAM potential for Nb produces the 
degenerate core. We determine the core-structure for a supercell containing 
about 900,000 atoms. The atoms are arranged in a cylindrical slab oriented such
that the $x$ axis is along the $[1 \bar2 1]$ direction, the $y$ axis is along 
$[\bar1 0 1]$, and the $z$ axis is along $[1 1 1]$. The radius of the cylinder 
is 60 nm. The supercell has 15 (111) planes in the $z$ direction. Periodic
boundary conditions are applied in the $z$ direction to simulate an 
infinitely-long straight screw dislocation. We insert a $(1/2)[1 1 1]$ screw 
dislocation into the ideal crystal by displacing all the atoms in the supercell
according to the dislocation's anisotropic elastic strain field~\cite{hirth82}.
The resulting structure provides an initial configuration for subsequent 
conjugate-gradient relaxation. Atoms that are less than $58$ nm from the center
of the cylinder are free to relax (the atomistic region) while the rest of the
atoms are fixed at their initial positions. This fixed boundary condition 
effectively extends the dimensions of the system to infinity in the $x$ and 
$y$ directions. There are no published DFT results for the $(1/2)\langle 111 
\rangle$ core-structure in Nb. F-S potentials~\cite{duesbery98} produce 
degenerate cores for the group VIB metals (Cr, Mo, W) and nondegenerate cores
for the group VB metals (V, Nb, Ta).

Duesbery and Vitek~\cite{duesbery98} propose a criterion that relates the 
$\{110\}$ $\gamma$ surface to the (1/2)$\langle 111 \rangle$ screw dislocation 
core structure. The criterion is based on results from F-S calculations and 
states that the degenerate core-structure forms if

\begin{equation}\label{eqn:GAMMA}
\gamma_{\{110\}}\left(b/3\right) < 2\gamma_{\{110\}}\left(b/6\right),
\end{equation}

\noindent
where $\gamma_{\{110\}}(b/3)$ and $\gamma_{\{110\}}(b/6)$ are the $\{110\}$
fault energies at $b/3$ and $b/6$ along $\langle 111 \rangle$, respectively. 
Our EAM potential produces $\gamma_{\{110\}}(b/3) = 0.033$ eV/{\AA}$^2$ and
$\gamma_{\{110\}}(b/6) = 0.014$ eV/{\AA}$^2$. These fault energies do not 
satisfy the criterion for the degenerate core, yet this is the core that our 
EAM potential favors. This suggests that the Duesbery-Vitek criterion is not 
generally valid, and that the shapes of the $\gamma$ surfaces and the type of 
core structure depend on the details of atomic interactions.

\begin{figure}
\includegraphics[width=0.45\textwidth]{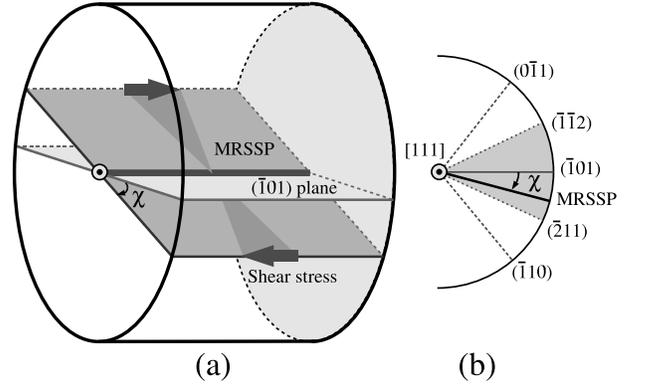}
\caption{\label{fig:MRSSP} Orientation of the maximum resolved shear stress 
plane (MRSSP) with respect to the $(\bar1 0 1)$ plane. (a) The angle between 
the MRSSP and the $(\bar1 0 1)$ plane is $\chi$. (b) Due to symmetry, we only 
need to consider $-30^\circ < \chi < +30^\circ$ (the shaded interval). This
range of angles includes the the $(\bar2 1 1)$, $(\bar1 0 1)$, and 
$(\bar1 \bar1 2)$ planes.}
\end{figure}

Figure~\ref{fig:screw} shows the relaxed core structure of the $(1/2)[111]$ 
screw dislocation, and its movement under pure shear stress acting parallel to 
the Burgers vector. We increase the strain on the crystal in small increments 
and allow the atoms in the atomistic region to relax after each increase in 
strain. The resulting shear stress acts in the maximum-resolved shear stress 
plane (MRSSP), and the dislocation moves when the stress reaches the 
critical-resolved shear stress (CRSS), i.e, the Peierls stress. We compute the 
CRSS for different orientations of the MRSSP. Figure~\ref{fig:MRSSP} shows that
the orientations of the MRSSP are defined by the angle $\chi$ the MRSSP makes 
with the $(\bar1 0 1)$ plane. It is sufficient to consider $-30^{\circ} < \chi 
< +30^{\circ}$ due to crystal symmetry. Figure~\ref{fig:screw}(b) shows that 
when the shear stress reaches the CRSS, the dislocation moves along the 
$(\bar1 \bar1 2)$ plane for all MRSSP orientations with $\chi < 25^{\circ}$. 
The net motion of the dislocation is in the $(\bar1 \bar1 2)$ plane. An
alternative way to view this motion is the dislocation moves along the 
$(\bar1 0 1)$ and $(0 \bar1 1)$ planes in steps of $(1/3)[1 \bar2 1]$ and 
$(1/3)[\bar2 1 1]$, producing an effective slip in the $(\bar1 \bar1 2)$ plane.
The same motion is observed in atomistic simulations of $(1/2)[111]$ screw 
dislocations in Ta using a F-S~\cite{ito01} potential and a model generalized 
pseudopotential theory potential~\cite{yang01}. Slip on $\{112\}$ and $\{110\}$
planes has been experimentally observed in Nb single crystals~\cite{duesbery66,
foxall67, kim09}.

\begin{figure}
\includegraphics[width=0.40\textwidth]{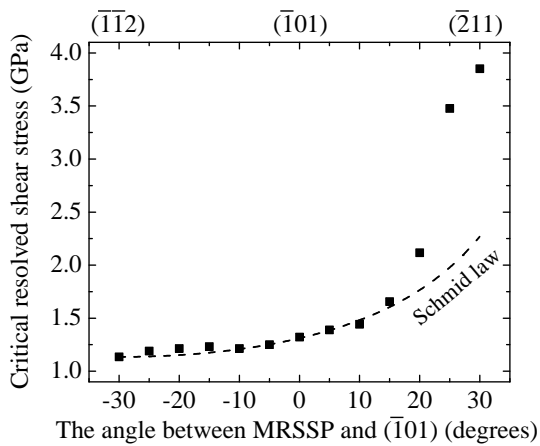}
\caption{\label{fig:peierls} The critical resolved shear stress (CRSS) for 
dislocation motion as a function of MRSSP orientation. The figure shows a 
deviation from the Schmid law when the angle between the MRSSP and the 
$(\bar1 0 1)$ plane is greater than about 15$^\circ$.}
\end{figure}

Figure~\ref{fig:peierls} shows the CRSS for various orientations of the MRSSP. 
The results clearly demonstrate the dependence of the CRSS on the sense of 
shearing and illustrates the well-known breakdown of the Schmid law in bcc 
metals~\cite{kubin82, christian83, duesbery89, vitek92, seeger95, pichl02, 
duesbery02, groger07}. This law assumes that components of the stress tensor 
other than shear in the slip plane in the slip direction play no role in the 
deformation process, and that the critical stress is independent of the sense 
of shearing. When $(\bar1 \bar1 2)$ is the slip plane, the Schmid-law 
dependence of the CRSS on $\chi$ has the form 1/cos($\chi$ + 30$^{\circ}$), 
drawn as a dashed curve in Fig.~\ref{fig:peierls}. Deviations from the Schmid
law becomes discernible for $\chi \gtrsim 15^{\circ}$, and rapidly increase as
the MRSSP approaches the $(\bar2 1 1)$ plane.

\subsection{\label{subsec:melting} Melting}

Our primary interest is solid-state simulations, but we also examine melting 
behavior. Morris {\it et al.}~\cite{morris07} state ``for EAM potentials, it 
has been commonly observed that the melting temperatures are significantly 
lower (30\% or more) than experimental values.'' Accordingly, an accurate 
melting temperature provides a challenging test for the potential. Two-phase 
melting simulations, in which the simulation cells contain solid and liquid in 
contact with each other, produce reliable melting temperatures. The 
liquid-solid interface provides nucleation sites for melting, thereby removing 
over-heating issues associated with single-phase melting simulations. Several 
methods based on this idea have been applied to the melting of metallic 
systems~\cite{morris94, belonoshko00, laio00, alfe02, morris02, morris07}. 

We follow the approach of Belonoshko {\it et al.}~\cite{belonoshko00}, in which
constant-$NPT$ MD simulations determine the melting temperature. Initially, 
half the simulation cell is liquid and the other half is bcc. For a given 
pressure, we compute the average volume for a series of simulations with 
increasing temperature. The liquid region of the simulation cell solidifies 
below the melting temperature, and the solid region liquefies above the melting
temperature. The volume of the system increases sharply across the melting 
temperature, indicating that a phase transition occurs. The average volume of 
each phase is constant at the melting point where the two phases coexist. Our 
simulation cells contain at least 16,500 atoms. Each MD simulation runs for 
5,000,000 steps with a 1 fs time step, and we average the volumes from the last
5,000 steps. We check the coexistence of the phases at the melting temperature 
using at least five independent simulations. We find that 130,000-atom 
simulations produce the same melting temperatures as 16,500-atom simulations.

\begin{figure}
\includegraphics[width=0.40\textwidth]{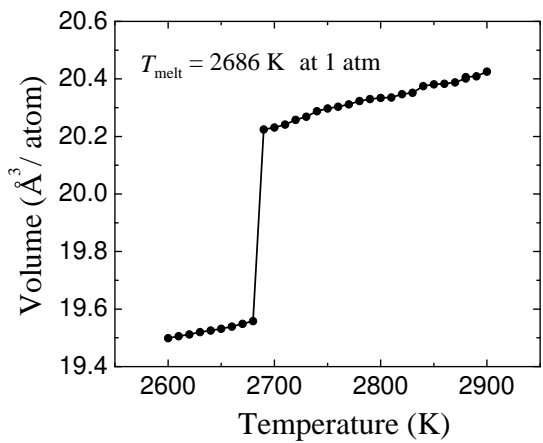}
\caption{\label{fig:melting} Two-phase melting simulations determine the 
melting temperature of our EAM potential. Initially, half the simulation cell 
contains liquid Nb and the other half contains bcc Nb. The liquid region of the
simulation cell solidifies below the melting temperature and the solid region 
melts above the melting temperature. The figure shows the equilibrium volume 
for each simulation temperature. There is a sharp jump in volume upon melting.}
\end{figure}

We compute melting temperatures for simulation cells containing liquid in 
contact with a $\{100\}$, $\{110\}$, or $\{111\}$ bcc surface. 
Figure~\ref{fig:melting} shows the increase in volume with temperature at 
$P$ = 1 atm for the liquid-$\{100\}$ interface. The melting temperatures at $P 
= 1$ atm from the liquid-$\{100\}$, liquid-$\{110\}$, and liquid-$\{111\}$ 
simulations are 2,686 K, 2,680 K, and 2,688 K, respectively. Each melting 
temperature has an error of $\pm 5$ K.  The average of the three melting 
temperatures is 2,685 K. The error between this value and the experimental 
melting temperature of 2,742 K is only 2\%. The agreement is excellent 
considering that the fitting database does not contain data from configurations
near the melting point.

\begin{figure}
\includegraphics[width=0.40\textwidth]{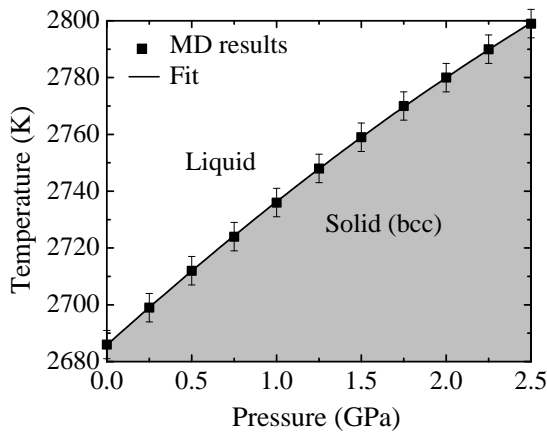}
\caption{\label{fig:meltingcurve} The melting curve of Nb. The points are EAM 
results and the solid line fits the values. The melting temperature is computed
for pressures to 2.5 GPa, with an error of $\pm 5$ K for each temperature. The 
EAM potential produces a melting temperature of 2,685 K at $P = 1$ atm. The 
experimental melting temperature at this pressure is 2,742 K. The error between
EAM and experiment is only 2\%. Melting temperatures have not been measured for
higher pressures.}
\end{figure}

We also determine the melting curve of Nb for pressures to 2.5 GPa. 
Figure~\ref{fig:meltingcurve} shows the increase in melting temperature with 
pressure. The points are results from constant-$NPT$ MD simulations, and the 
solid line is a quadratic fit through the values: $T = T_0 + \alpha P + \beta 
P^2$, where $T_0 = 2,685.8 \pm 0.2$ K, $\alpha = 53.9 \pm 0.3$ K/GPa, and 
$\beta = -3.4 \pm 0.1$ K/GPa$^2$. Each data point has an error of $\pm 5$ K. 
The melting curve of Nb has not been measured.

\begin{figure}
\includegraphics[width=0.40\textwidth]{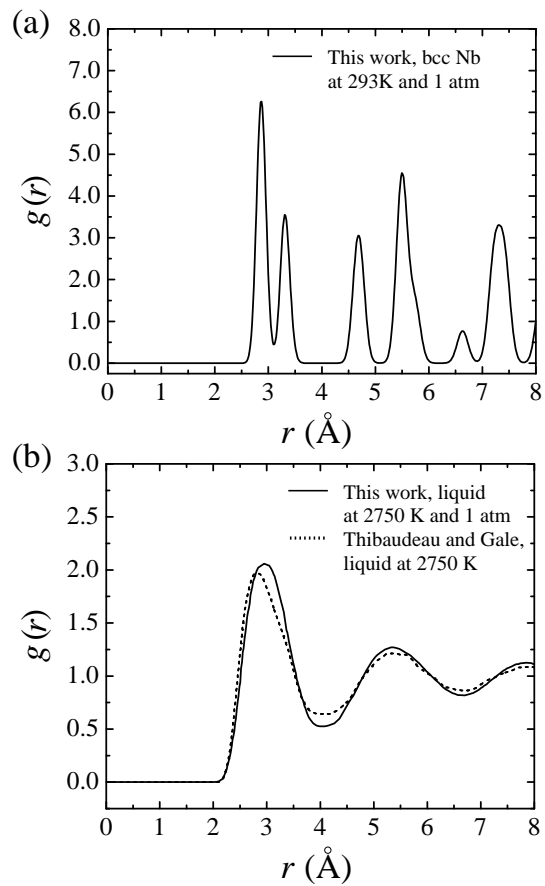}
\caption{\label{fig:RDF} Radial distribution functions (RDF). (a) Nb is bcc at 
293 K. (b) Liquid Nb at 2,750 K. No experimental data is available, so we 
compare the liquid result to the RDF from an EAM potential intended for 
simulating liquid Nb~\cite{thibaudeau08}. The two potentials produce similar 
results. The first and second neighbor shells in the solid merge into a single 
peak in the liquid, with similar behavior for higher-order neighbor shells.}
\end{figure}

Figure~\ref{fig:RDF} shows the radial distribution functions (RDF) for bcc Nb 
at 273 K and 1 atm, and liquid Nb at 2,750 K and 1 atm. We determine the RDFs 
by averaging position data from over 1,000 MD simulation steps. No experimental
data is available for liquid Nb, so we compare our prediction of the liquid RDF
to the result from an EAM potential intended for simulating liquid 
Nb~\cite{thibaudeau08}. Both potentials predict that groups of bcc peaks merge 
to form wider peaks in the liquid but there are small differences in the 
maxima of the peaks.

\section{\label{sec:conclude} Summary}

We construct an accurate and reliable EAM potential for Nb as the first step in
alloy potential development. The force-matching program {\sc potfit} optimizes 
the EAM functions to a database of well-converged DFT forces, energies, and 
stresses. The potential accurately reproduces properties tied to the fitting 
data, and shows excellent agreement with DFT and experiment for a large number 
of other quantities that are related to configurations not included in the 
fitting database. The potential describes structural and elastic properties, 
defects, and thermodynamic behavior. While the potential may not be well suited
for shock-wave or radiation damage studies, it performs very well in all other 
situations we have tested. The potential also serves as a viable starting point
for constructing accurate EAM potentials for Nb alloys.

\begin{acknowledgments}

We thank Richard G. Hennig for providing the DFT phonon results. We thank 
Richard G. Hennig, Thomas J. Lenosky, Dallas R. Trinkle, and Murray S. Daw for 
useful discussions. This work was supported by DOE-Basic Energy Sciences, 
Division of Materials Sciences (DE-FG02-99ER45795). Computational resources 
were provided in part by an allocation of computing time from the Ohio 
Supercomputer Center. This research also used resources of the National Energy 
Research Scientific Computing Center, which is supported by the Office of 
Science of the U.S. Department of Energy under Contract No. DE-AC02-05CH11231.
\end{acknowledgments}

\appendix*
\section{Function modifications}

In this appendix we discuss modifications to $\phi(r)$ and $\rho(r)$ for small 
$r$, and to $F(n)$ for small and large $n$. In MD simulations, fluctuations 
can move atoms closer together than the minimum interatomic distance in the 
fitting database. The {\sc potfit} program accounts for this by extending 
$\phi$ and $\rho$ to $r$ values smaller than the inner cutoff radius. The cubic
polynomials in the range $2.073 < r < 2.240 \; {\rm \AA}$ are extended down to 
$1.738 < r < 2.240 \; {\rm \AA}$. Likewise, {\it potfit} extends $F$ to $n$ 
values smaller than the inner cutoff, and $n$ values larger than the outer 
cutoff. The cubic polynomial in the range $0.0775 < n < 0.209$ is extended down
to $-0.0186 < n < 0.209$, and a very steep cubic polynomial is added for 
$1.000 < n < 1.264$. Requiring continuity of $F$ and its first and second 
derivatives at $n = 1.000$ determines three coefficients of the steep cubic 
function. Setting $F$ equal to 4.828 eV at $n = 1.264$ determines the final 
coefficient.

Despite these modifications, the potential is not repulsive enough for 
high-temperature and high-pressure simulations, where atoms closely approach 
one another. To overcome this limitation, we modify $\phi$ for 
$1.738 < r < 2.073 \; {\rm \AA}$. We replace the function in this range by a 
steeper cubic polynomial. The continuity of $\phi$ and its first and second 
derivatives at $r = 2.073 \; {\rm \AA}$ determines three coefficients. 
Setting the first derivative at $r = 1.738 \; {\rm \AA}$ equal to four times 
the first derivative at $r = 2.073 \; {\rm \AA}$ determines the final 
coefficient. This modified potential is repulsive enough at small atomic 
separations to prevent collapse problems for all temperature and pressure 
ranges we investigated. 

We also modify the extension of $F$ for small $n$ values to properly describe 
the cohesive energy. The minimum of the EAM energy per atom versus volume curve
for bcc Nb equals the cohesive energy, but the embedding energy is not zero for
$n = 0$ when the atoms are far apart. Therefore, we replace the {\sc potfit} 
modification for small $n$ by a different cubic polynomial. We choose three 
coefficients to ensure continuity of the embedding function and its first 
and second derivatives at $n = 0.0775$. We determine the final coefficient 
by setting $F(0) = 0$. Table~\ref{tab:knots} lists the optimized spline knots, 
boundary conditions, and modifications of $\phi$, $\rho$, and $F$.


\begin{thebibliography}{93}
\bibitem[{BMn()}]{BMnote}
\bibinfo{note}{The third-order Birch-Murnaghan equation of state is $E_{\rm
  BM}(V) = E_0 + (9/8) B V_0 \left[ (V_0/V)^{2/3} -1 \right]^2 + (9/16) B V_0
  (B'-4) \left[ (V_0/V)^{2/3} -1 \right]^3$, where $E_0$ is the minimum of the
  energy vs. volume curve, $B$ is the bulk modulus, $V_0$ is the equilibrium
  volume, and $B'$ is the pressure derivative of $B$.}
\end{thebibliography}

\end{document}